\documentclass[11pt]{article}
\usepackage{hyperref}
\usepackage{amsmath}
\usepackage{amssymb}
\usepackage{bm}
\usepackage{latexsym}
\begin{document}
\title{On the perturbation theory in spatially closed background}
\author{Ali A. Asgari \\Amir H. Abbassi*\\ Jafar Khodagholizadeh\\
\\\vspace{6pt} Department of Physics, School of Sciences,\\ Tarbiat Modares University, P.O.Box 14155-4838, Tehran, Iran\\}
\maketitle
\begin{abstract}
In this article, we investigate some features of the perturbation theory in a spatially closed universe. We will show that the perturbative field equations in a spatially closed universe always have two independent adiabatic solutions provided that the wavelengths of perturbation modes are very longer than the Hubble horizon. It will be revealed that these adiabatic solutions don’t depend on the curvature directly. We also propound a new interpretation for the curvature perturbation in terms of the unperturbed background geometry.   
\end{abstract}
\section{Introduction}
\label{intro}
The theory of the linear perturbations is an important part of the modern cosmology which explains CMB anisotropies and structure formation origin. This theory has been investigated for a spatially flat universe exceedingly\cite{Ref1,Ref2,Ref3,Ref4,Ref5,Ref6,Ref7,Ref8,Ref9,Ref10}. However, observational data points out a universe with $\Omega_{\Lambda}\cong.68$\cite{Ref11}. Existence of a positive cosmological constant necessitates a de Sitter spacetime for the vacuum background. From the different forms of the de Sitter spacetime with $K=0,\pm1 $, merely $ K=1 $ case, namely, \textit{Lorentzian de Sitter spacetime} is maximally symmetric, maximally extended and also geodesically complete\cite{Ref12}. So in the following we assume $ \Lambda>0$ and $ K=1 $  for the vacuum background. Furthermore, it seems hard to believe that the total density of the universe has been tuned in $ \rho _{{crit}_ 0} $ exactly, because despite the fact that the observational data indicate $ \Omega_K =0 $ \cite{Ref11}, this fine-tuning seems somehow unlikely. Moreover, if $ \Omega_{tot} $ equals to $ +1 $ exactly, this cannot last forever because of the instability \cite{Ref13}. On the other hand, there are some reasons that the universe may have positive spatially curvature with non-trivial topology. In other words, some positive curvature models with non-trivial topology can solve the problem of the CMB quadrupole and octopole suppression and also mystery of the missing fluctuations which appear in the concordance model of cosmology \cite{Ref14,Ref15,Ref16,Ref17,Ref18}. So these reasons augment the probability of spatially closed case and it seems necessary to investigate the theory of small fluctuations in spatially closed universes.\\ The outline of this article is as follows. In Section 2 we derive the equations governing the linear perturbations in FLRW universe without fixing $ K $. In Section 3 we study the spectral and stochastic properties of these perturbations for the case $ K=1 $ and in Section 4 the gauge problem will be discussed. Finally, in the last section we derive two independent adiabatic solutions for the obtained equations with $ K=1 $ while the perturbations scales go outside of the Hubble horizon. It will be seen that one of these solutions is decaying, so it has not cosmological significance. We also deduce a new geometrical interpretation for the curvature perturbation as the \textit{conformal factor} of the spatial section of the background spacetime. Furthermore, we will show that for the supper-Hubble scales, curvature has no direct effect on the universe evolution.

\section{The perturbed spacetime}
\label{sec2}
We assume during most of the time the departures from homogeneity and isotropy have been very small, so that they can be treated as the first order perturbations. The total perturbed metric is 
\begin{equation}
g_{\mu \nu }  = \bar g_{\mu \nu }  + h_{\mu \nu } ,
\end{equation}
where $  \bar g_{\mu \nu } $  and $ h_{\mu \nu }  $ are the unperturbed metric and the first order perturbation respectively. Note that  $  \bar g_{\mu \nu } $ is the FLRW metric which in the comoving quasi-Cartesian coordinates can be written as\cite{Ref2}
\begin{align*}
 &g_{00}  =  - 1 ,\quad g_{0i}  = g_{i0}  = 0 ,\\&
 g_{ij}  = a^2 \left( t \right)\tilde g_{ij}=a^2 \left( t \right)\left( \delta _{ij}  + K\frac{{x^i x^j }}{{1 - K{\bf{x}}^2 }}\right)    ,
\end{align*}
 Bar over any quantity denotes its unperturbed value. Perturbing the metric leads to perturbing the connection and Ricci tensor as\cite{Ref2}
\begin{equation}
\delta \Gamma _{\mu \nu }^\lambda   = \frac{1}{2}\bar g^{\lambda \rho } \left( { - 2h_{\rho \eta } \bar \Gamma _{\mu \nu }^\eta   + \partial _\mu  h_{\nu \rho }  + \partial _\nu  h_{\mu \rho }  - \partial _\rho  h_{\mu \nu } } \right) ,
\end{equation}
and
\begin{equation}
\delta R_{\mu \nu }= \partial _\nu  \delta \Gamma _{\mu \lambda }^\lambda   - \partial _\lambda  \delta \Gamma _{\mu \nu }^\lambda   + \delta \Gamma _{\mu \rho }^\lambda  \bar \Gamma _{\nu \lambda }^\rho   + \delta \Gamma _{\nu \rho }^\lambda  \bar \Gamma _{\mu \lambda }^\rho  - \delta \Gamma _{\mu \nu }^\lambda  \bar \Gamma _{\rho \lambda }^\rho - \delta \Gamma _{\lambda \rho }^\lambda  \bar \Gamma _{\mu \nu }^\rho . 
\end{equation}
The perturbative form of the Einstein's field equations may be written as 
\begin{equation}\label{c}
\delta R_{\mu \nu }  =  - 8\pi G\delta S_{\mu \nu } , 
\end{equation}
where 
\begin{equation}
\delta S_{\mu \nu }  = \delta T_{\mu \nu }  - \frac{1}{2}\bar g_{\mu \nu } \delta T - \frac{1}{2}\bar Th_{\mu \nu }, 
\end{equation}
On the other hand, the perturbation of the energy-momentum conservation law gives
\begin{equation}\label{d}
\partial _\mu  \delta T^\mu\smallskip _ \nu   - \delta \Gamma _{\mu \nu }^\lambda  \bar T^\mu \smallskip _{{\rm{  }}\lambda }  - \bar \Gamma _{\mu \nu }^\lambda  \delta T^\mu \smallskip _\lambda   + \bar \Gamma _{\mu \lambda }^\mu  \delta T^\lambda \smallskip _\nu   + \delta \Gamma _{\mu \lambda }^\mu  \bar T^\lambda \smallskip _{{\rm{  }}\nu }  = 0 .
\end{equation}
Setting $ \nu $ equal to 0 and $ i $ gives equations of energy and momentum conservation respectively. The explicit form of these equations is too lengthy and complicated, so we avoid expressing them here. Fortunately there is a mathematical technique, which simplifies these equations remarkably \cite{Ref3,Ref4,Ref5}. According to this technique we can decompose $ h_{\mu\nu} $ into four scalars, two divergenceless, spatial vector and a symmetric, traceless,  divergenceless spatial tensor as follows 
\begin{align}
&h_{00}  =  - E\label{i} ,\\&
h_{i0}  = a\left( {\nabla _i F + G_i } \right)\label{j} ,\\&
h_{ij}  = a^2 \left( {A\tilde g_{ij}  + H_{ij} B + \nabla _i C_j  + \nabla _j C_i  + D_{ij} } \right)\label{k} ,
\end{align}
where $ \nabla_{i} $ is the covariant derivative respect to the spatial unperturbed metric $\bar g_{ij}(=a^2\tilde g_{ij})$ and $ H_{ij}=\nabla _i \nabla _j $ is the \textit{covariant Hessian operator} . All the perturbations $A, B, E, F, C_{i}, G_{i}$ and $D_{ij}$ are functions of $t$ and $\bold x$ which satisfy 
\begin{align}
 &\nabla ^i C_i  = \nabla ^i G_i  = 0\label{l} , \\& 
 \tilde g^{ij} D_{ij}  = 0,\quad \nabla ^i D_{ij}  = 0,\quad D_{ij}  = D_{ji}\label{m} . 
 \end{align}
Eq.(\ref{j}) is generalization of the Helmholtz's decomposition theorem from $\mathbb{R}^3$ to the Riemannian manifolds. Eq.(\ref{k}) is also a theorem in Riemannian geometry\cite{Ref19,Ref20}. According to this theorem, every rank 2 symmetric tensor on a \textit{compact} Riemannian manifold can be uniquely represented in the Eq.(\ref{k}) form. It is possible to carry out a similar decomposition of the energy- momentum tensor. One can show that \cite{Ref2}
\begin{align*}
 &\delta T_{00}  =  - \bar \rho h_{00}  + \delta \rho , \nonumber \\& 
 \delta T_{i0}  = \bar ph_{i0}  - \left( {\bar \rho  + \bar p} \right)\delta u_i , \nonumber \\&  
 \delta T_{ij}  = \bar ph_{ij}  + a^2 \tilde g_{ij} \delta p . \nonumber 
  \end{align*}
 We can decompose velocity perturbation $ \delta u_i $ into the gradient of a scalar (velocity potential) $ \delta u $ and a transverse vector $ \delta u^V_i $
\begin{equation}\label{a}
\delta u_i=\nabla _i (\delta u)+\delta u^V_i , \quad \nabla^i \delta u^V _i=0 
\end{equation} 
We may consider imperfectness of the cosmic fluid by adding a term $ \Pi _{ij} $  to $\delta T_{ij} $. $ \Pi _{ij} $ is known as \textit{anisotropic inertia} tensor field of the fluid and may be decomposed just like as $ h_{ij} $
\begin{equation}\label{n}
\Pi _{ij}  = a^2 \left( {H_{ij} \Pi ^S  + \nabla _i \Pi _j^V + \nabla _j \Pi _i^V + \Pi _{ij}^T } \right) ,
\end{equation}
where $ \Pi^V_i $ and $ \Pi^T_{ij} $ satisfy conditions analogous to the Eqs.(\ref{l}) and (\ref{m}), which satisfied by $ C_i $ and $ D_{ij} $ in return.
In Eq.(\ref{n}) there is no term proportional to $ \tilde g_{ij} $, because $\delta T_{ij} $ itself contains such term. Finally, we have
\begin{align}
& \delta T_{00}  =  - \bar \rho h_{00}  + \delta \rho \label{o}, \\ &
 \delta T_{i0}  = \bar ph_{i0}  - \left( {\bar \rho  + \bar p} \right)\left( {\nabla _i \delta u + \delta u_i^V } \right) \label{p} ,\\&
\delta T_{ij} = \bar ph_{ij}  + a^2 \big( \tilde g_{ij} \delta p + H_{ij} \Pi ^S  + \nabla _i \Pi _j^V  \nonumber\\&
 \qquad+ \nabla _j \Pi _i^V  + \Pi _{ij}^T  \big)\label{q} .
 \end{align}
Now let's define \textit{Laplace-Beltrami operator} 
\begin{equation*}
\nabla ^2  = \bar g^{ij}H_{ij}=\bar g^{ij} \nabla _i \nabla _j .
\end{equation*}
Thus, for scalar field $\mathcal{S}$ we have
\begin{equation}\label{r}
a^2 \nabla ^2\mathcal{S} = \tilde g^{ij} \partial _i \partial _j \mathcal{S} - 3K\left( {\partial _i \mathcal{S}} \right)x^i . 
\end{equation}
Also for vector field $ \mathcal{V}_i $ and tensor field $\mathcal{T}_{ij} $ we can write
\begin{equation}\label{s}
a^2 \nabla ^2 \mathcal{V}_i  = \tilde g^{jk} \partial _j \partial _k \mathcal{V}_i  - K\mathcal{V}_i  - 2K\left( {\partial _i \mathcal{V}_j } \right)x^j  - 3K\left( {\partial _j \mathcal{V}_i } \right)x^j ,
\end{equation}
\begin{align}\label{t}
 a^2 \nabla ^2 \mathcal{T}_{ij}  &= \tilde g^{kl} \partial _k \partial _l \mathcal{T}_{ij}  - 2K\mathcal{T}_{ij}
- 2K\left( {\partial _i \mathcal{T}_{jk} } \right)x^k- 2K\left( {\partial _j \mathcal{T}_{ik} } \right)x^k \nonumber \\&
 - 3K\left( {\partial _k \mathcal{T}_{ij} } \right)x^k+ 2K^2 \tilde g_{ij} \mathcal{T}_{kl} x^k x^l .
\end{align}
Substituting the Eqs.(\ref{i}), (\ref{j}), (\ref{k}), (\ref{a}), (\ref{o}), (\ref{p}) and (\ref{q}) in the field and conservation equations namely Eqs.(\ref{c}) and (\ref{d}) and also separating the terms containing $\tilde g_{ij} $ ,${\nabla _i }$ and $H_{ij}  $, accompanied by using Eqs.(\ref{r}), (\ref{s}) and (\ref{t}) results in three idependent sets of coupled equations
\subsection{Scalar mode equations}
These equations involve just scalars:\\
\begin{align}
 & 2KA + \dot aa^2 \nabla ^2 F - 3a\dot a\dot A - \frac{1}{2}\dot aa^3 \nabla ^2 \dot B - \frac{1}{2}a^2 \ddot A + \left( {2\dot a^2  + a\ddot a} \right)E +\frac{1}{2} a\dot{a} \dot{E}  \nonumber \\ &
 + \frac{1}{2}a^2 \nabla ^2 A= 4\pi Ga^2 \left( { - \delta \rho + \delta p + a^2 \nabla ^2 \Pi ^S } \right), \\ \nonumber \\& 
 4\dot aF - 3a\dot a\dot B + 2a\dot F - a^2 \ddot B + E + A= - 16\pi Ga^2 \Pi ^S , \\ \nonumber\\ & 
 a\dot A - \dot aE - Ka\dot B + 2KF = 8\pi Ga\left( {\bar \rho  + \bar p} \right)\delta u, \\ \nonumber \\ & 
 3\frac{{\dot a}}{a}\dot A + a\dot a\nabla ^2 \dot B + \frac{3}{2}\ddot A + \frac{1}{2}a^2 \nabla ^2 \ddot B - \frac{3}{2}\frac{{\dot a}}{a}\dot E - \dot a\nabla ^2 F - a\nabla ^2 \dot F - \frac{1}{2}\nabla ^2 E \nonumber \\ & 
- 3\frac{{\ddot a}}{a}E=  - 4\pi G\left( {\delta \rho  + 3\delta p + a^2 \nabla ^2 \Pi ^S } \right),\\ \nonumber\\ &
 \frac{{\partial \delta \rho }}{{\partial t}} + \nabla ^2 \left[ { - a\left( {\bar \rho  + \bar p} \right)F + \left( {\bar \rho  + \bar p} \right)\delta u + a\dot a\Pi ^S } \right] + \frac{1}{2}\left( {\bar \rho  + \bar p} \right)\left( {3\dot A + a^2 \nabla ^2 \dot B} \right)  \nonumber\\&
 + 3\frac{{\dot a}}{a}\left( {\delta \rho  + \delta p} \right) = 0\label{g},\\ \nonumber\\& 
\mathop {\bar p}\limits^.  \delta u + \left( {\bar \rho  + \bar p} \right)\frac{{\partial \delta u}}{{\partial t}} + \frac{1}{2}\left( {\bar \rho  + \bar p} \right)E + \delta p + a^2 \nabla ^2 \Pi ^S+ 2K\Pi ^S  = 0.
 \end{align}
\subsection{Vector mode equations}
\begin{equation}
2\dot aG_i  - 3a\dot a\dot C_i  + a\dot G_i  - a^2 \ddot C_i  =  - 16\pi Ga^2 \Pi _i^V,
\end{equation}
\begin{align}
&- \frac{1}{2}a^3 \nabla ^2 \dot C_i  + \frac{1}{2}a^2 \nabla ^2 G_i  - Ka\dot C_i  + KG_i = 8\pi Ga\left( {\bar \rho  + \bar p} \right)\delta u_i^V  ,\\ \nonumber\\ &
\mathop {\bar p}\limits^. \delta u_i^V  + \left( {\bar \rho  + \bar p} \right)\frac{{\partial \delta u_i^V }}{{\partial t}} + a^2 \nabla ^2 \Pi _i^V  + 2K\Pi _i^V  = 0.
 \end{align}
 \subsection{Tensor mode equation}
 \begin{equation}
 a^2 \nabla ^2 D_{ij}  - 3a\dot a\dot D_{ij}  - a^2 \ddot D_{ij}  - 2KD_{ij}  =  - 16\pi Ga^2 \Pi _{ij}^T.
 \end{equation}
As previously mentioned, in the linear perturbation theory, the scalar, vector and tensor modes evolve independently. The vector and tensor modes are not important for structure formation because they produce no  density perturbation, albeit they affect on the CMB anisotropy. 
\section{Fourier decomposition and random fields}
\label{sec3}
In this section, we study the spectral and stochastic properties of the perturbations for the case $ K=+1 $. Albeit the equations have been derived in Section \ref{sec2} describe the time evolution of the perturbative quantities, viewed as functions of position (at fixed time) they are considered as random fields on $ S^3(a) $, because they are defined on a homogeneous and isotropic space \cite{Ref7,Ref21}. Now we investigate the stochastic properties of perturbations for every mode separately.
\subsection{Scalar perturbations and scalar random fields}
An important class of random fields are described by their Fourier transformations. There are many different Fourier transform conventions, however here our intention is the expansion of each mode of the perturbation fields in terms of the corresponding eigenfunctions of the Laplace-Beltrami operator. Thus, we have to find the eigen functions of $ \nabla^2 $ on $ S^3(a) $. For scalar mode we have
\begin{equation}\label{u}
\nabla ^2 \Phi  = \Xi \Phi , 
\end{equation}
where $\nabla ^2  = \bar g^{ij} H_{ij} $. In pseudo-spherical coordinates with the line element
\begin{align}
 ds^2=a^2(d\chi^2+{\sin ^2 \chi }d\theta^2+{\sin ^2 \chi }{\sin ^2 \theta } d\varphi^2), 
\end{align}
the Eq.(\ref{u}) gives
\begin{align}\label{v}
\frac{1}{{a^2 }}\bigg( \frac{{\partial ^2 \Phi }}{{\partial \chi ^2 }} + \frac{1}{{\sin ^2 \chi }}\frac{{\partial ^2 \Phi }}{{\partial \theta ^2 }} + \frac{1}{{\sin ^2 \chi \sin ^2 \theta }}\frac{{\partial ^2 \Phi }}{{\partial \varphi ^2 }}+ 2\cot \chi \frac{{\partial \Phi }}{{\partial \chi }} + \frac{{\cot \theta }}{{\sin ^2 \chi }}\frac{{\partial \Phi }}{{\partial \theta }} \bigg) = \Xi \Phi .
\end{align}
Solving Eq.(\ref{v}) gets to the following eigenvalues and eigenfunctions \cite{Ref22,Ref23,Ref24,Ref25}
\begin{align}
&\Xi= \Xi _n  = \frac{{1 - n^2 }}{{a^2 }},\quad n = 1,2,...\\&
\Phi  =\mathcal{Y} _{nlm} \left( {\chi ,\theta ,\varphi } \right) =\Pi  _{nl} \left( \chi  \right)Y_{lm} \left( {\theta ,\varphi } \right),\\&
 n = 1,2,...,\quad l \le n - 1,\quad \left| m \right| \le l \nonumber,
\end{align}
where
\begin{equation}
\Pi _{nl} \left( \chi  \right) = \frac{{\left( {2l} \right)!!}}{{\sqrt{a^3}}}
\sqrt {\frac{2}{\pi }\frac{{n\left( {n - l - 1} \right)!}}{{\left( {n + l} \right)!}}} \sin ^l \chi C_{n - l - 1}^{l + 1} \left( {\cos \chi } \right) ,
\end{equation}
are known as\textit{ Fock harmonics} \cite{Ref22,Ref25}. Also $Y_{lm} $ and $C_n^\lambda $ are scalar spherical harmonics on $ S^2 $ and Gegenbauer (ultraspherical) polynomials respectively. It can be shown that
\begin{equation}
\int\limits_{S^3 \left( a \right)} {d\mu \mathcal{Y} _{nlm}  \left( {\chi ,\theta ,\varphi } \right)} \mathcal{Y} ^* _{n'l'm'} \left( {\chi ,\theta ,\varphi } \right) = \delta _{nn'} \delta _{ll'} \delta _{mm'} ,
\end{equation}
where $ d\mu=a^3 \sin^2\chi \sin\theta d\chi d\theta d\varphi  $ is the invariant volume element on $ S^3(a) $. Scalar harmonics on $ S^3(a) $ also can be expressed in terms of \textit{Jacobi polynomials} or associated Legendre functions \cite{Ref26,Ref27}. Furthermore $ \mathcal{Y} _{nlm} $ s constitute a complete orthonormal set for expansion of any scalar field on  $ S^3(a) $. Thus, for scalar perturbative quantity $ A(t,\textbf{x}) $ at some instant (which thereafter will be denoted by$ A(\textbf{x}) $) we can write
\begin{equation}
A\left(\textbf{x}  \right) = \sum\limits_{nlm} {A_{nlm} \mathcal{Y}_{nlm} } \left( {\chi ,\theta ,\varphi } \right) .
\end{equation}
$A_{nlm}  $ just like $ A\left( {\bf{x}} \right) $ is a scalar random field. Apart from distribution function of $A_{nlm}  $, its simplest statistics are mean value and two-point covariance function, the latter is defined by $ \langle A_{nlm} A^* _{n'l'm'}  \rangle $. Here $ \langle\quad \rangle $ means ensemble average which equals to spatial average according to the ergodic theorem \cite{Ref7}. \\
The homogeneity of $S^3(a)$ implies for any pair of scalar random field $A$ and $B$
\begin{equation}
\langle A\left( {\bf{x}} \right)B^* \left( {{\bf{x'}}} \right) \rangle  =  \langle A\left( {{\bf{x}} + {\bf{R}}} \right)B^* \left( {{\bf{x'}} + {\bf{R}}} \right) \rangle .
\end{equation}
($\textbf{R}$ is an arbitrary 3-vector in $\mathbb{R}^3$) Thus $ \langle A\left( {\bf{x}} \right)B^* \left( {{\bf{x'}}} \right) \rangle $ must be just function of ${\bf{x}} - {\bf{x'}} $. This implies that
\begin{equation}\label{w}
\langle A_{nlm} A^* _{n'l'm'}  \rangle  \propto \delta _{nn'} \delta _{ll'} \delta _{mm'} .
\end{equation}
 It means $A_{nlm} $ and $  A_{n'l'm'} $ are uncorrelated random variables for different indices (indeed it results from the homogeneity of the spatial section of the backgrond spacetime). The homogeneity also implies that the coefficient of proportionality in Eq.(\ref{w}) is just function of $ n $ i.e.
\begin{equation}
 \langle A_{nlm} A^* _{n'l'm'}  \rangle  = P_A^0 \left( n \right)\delta _{nn'} \delta _{ll'} \delta _{mm'} .
\end{equation}
$ P_A^0 \left( n \right) $ is power spectrum or spectral density of $ A $ (the superscript "0" over $ P $ states corresponding spin of the random field) which depends on distribution function governing on $ A $. Moreover we have 
\begin{equation}\label{x}
 \langle A_{nlm} B^* _{n'l'm'} \rangle  = P_{A,B}^0 \left( n \right)\delta _{nn'} \delta _{ll'} \delta _{mm'} ,
\end{equation}
which $ P_{A,B}^0 \left( n \right) $  is joint power  spectrum of $A $ and $B $ \cite{Ref28,Ref29}.
One may define the correlation coefficient between $ A $ and$ B $:
\begin{equation}
\Delta _{A,B} \left( n \right) = \frac{{P_{A,B}^0 \left( n \right)}}{{\sqrt {P_A^0 \left( n \right)P_B^0 \left( n \right)} }} .
\end{equation}
$ - 1 \le \Delta _{A,B} \left( n \right) \le 1$ and two extreme values $\Delta _{A,B} \left( n \right) =  + 1 $ and $\Delta _{A,B} \left( n \right) =  - 1 $ correspond respectively to full correlation and full anti-correlation \cite{Ref29}.\\
Finally, let's define spectral index of random field $A$ as
\begin{equation}
\mathfrak{N}_A  = 4 + \frac{n}{{P_A^0 \left( n \right)}}\frac{{dP_A^0 \left( n \right)}}{{dn}}.
\end{equation}
Now we prove that the homogeneity of the universe yields Eq.(\ref{x}). At first, let's calculate $\langle A\left( {\bf{x}} \right)B^* \left( {{\bf{x'}}} \right) \rangle  $
\begin{align}
\langle A \left( {\bf{x}} \right)B^* \left( {{\bf{x'}}} \right) \rangle &= \sum\limits_{nlm} {\sum\limits_{n'l'm'} { \langle A_{nlm} B^* _{n'l'm'} } }  \rangle
 \mathcal{Y}_{nlm} \left( {\bf{x}} \right)\mathcal{Y}_{n'l'm'}^* \left( {{\bf{x'}}} \right) \nonumber \\& 
=\sum\limits_{nlm} {P_{A,B}^0 } \left( n \right)\mathcal{Y}_{nlm} \left( {\bf{x}} \right)\mathcal{Y}_{n'l'm'}^* \left( {{\bf{x'}}} \right)\notag \\&
= \sum\limits_{nl} {\frac{{2l + 1}}{{4\pi }}} P_{A,B}^0 \left( n \right)\Pi _{nl} \left( \chi  \right)\Pi _{nl} \left( {\chi '} \right)P_l \left( {{\bf{\hat x}}.{\bf{\hat x'}}} \right).
\end{align}
On the other hand, according to the addition formula of Gegenbauer polynomials (Fock harmonics)\cite{Ref30} we have
\begin{equation}
\frac{{\sin n\gamma }}{{\sin \gamma }} = \frac{\pi }{2}\frac{{a^3 }}{n}\sum\limits_{l=0}^{n-1} {\left( {2l + 1} \right)} \Pi _{nl} \left( \chi  \right)\Pi _{nl} \left( {\chi '} \right)P_l \left( {{\bf{\hat x}}.{\bf{\hat x'}}} \right) ,
\end{equation}
where $ \cos \gamma  = \cos \chi \cos \chi ' + \sin \chi \sin \chi '\left( {{\bf{\hat x}}.{\bf{\hat x'}}} \right) $.
Consequently 
\begin{equation}\label{y}
\langle A\left( {\bf{x}} \right)B^* \left( {{\bf{x'}}} \right) \rangle  = \frac{1}{{2\pi ^2 a^3 }}\frac{1}{{\sin \gamma }}\sum\limits_{n = 1}^\infty  n P_{A,B}^0 \left( n \right)\sin n\gamma ,
\end{equation}
which is invariant under following transformations obviously
\begin{equation}
\left\{ {\begin{array}{*{20}c}
   {\varphi  \to \varphi  + \delta }  \\
   {\varphi ' \to \varphi ' + \delta }  \\
\end{array}},\quad \right.\left\{ {\begin{array}{*{20}c}
   {\varphi  = \varphi '}  \\
   {\theta  \to \theta  + \delta }  \\
   {\theta ' \to \theta ' + \delta }  \\
\end{array}},\quad \right.\left\{ {\begin{array}{*{20}c}
   {\varphi  = \varphi '}  \\
   {\theta  = \theta '}  \\
   {\chi  \to \chi  + \delta }  \\
   {\chi ' \to \chi ' + \delta }  \\
\end{array}} \right.
\end{equation}
Moreover one can show 
\begin{equation}
\cos \gamma  = 1 - \frac{1}{2}\left( {\cos \chi  - \cos \chi '} \right)^2  - \frac{1}{2}\left| {{\bf{x - x'}}} \right|^2 
\end{equation}
This shows for $ \chi  = \chi ' $, $ \cos\gamma $ is a function of $\left| {{\bf{x - x'}}} \right| $, thus we conclude that Eq.(\ref{x}) depends merely on $\left| {{\bf{x - x'}}} \right| $.
Now let's turn to the vector mode.
\subsection{Vector perturbations and vector random fields}
In order to investigate vector perturbation we should find vector spherical harmonics on $S^3(a)$ at first. They are solutions of the following equation
\begin{equation}\label{y}
\nabla ^2 V_i  = \Upsilon V_i,\quad \nabla ^i V_i  = 0 
\end{equation}
The transversality condition is added as a constraint, because every vector perturbation in cosmology $ (C_i, G_i,$    $ \Pi^V_i) $ is divergenceless. It can be shown that the vector spectrum of $ S^3(a)$ is \cite{Ref22,Ref23,Ref24,Ref25}
\begin{equation}
\Upsilon  = \Upsilon _n  = \frac{{2 - n^2 }}{{a^2 }}.\quad n = 2,3,...
\end{equation}
and there are two independent eigenfunctions which in pseudo-spherical coordinates are
\begin{align}
 &\left( {V_1^o } \right)_{nlm}  = 0, \nonumber\\ &
 \left( {V_2^o } \right)_{nlm}  =  - \frac{a}{{\sqrt {l\left( {l + 1} \right)} }}\sin \chi \Pi _{nl} \left( \chi  \right)\frac{1}{{\sin \theta }}\frac{{\partial Y_{lm} }}{{\partial \varphi }}, \nonumber\\ & 
 \left( {V_3^o } \right)_{nlm}  = \frac{a}{{\sqrt {l\left( {l + 1} \right)} }}\sin \chi \Pi _{nl} \left( \chi  \right)\sin \theta \frac{{\partial Y_{lm} }}{{\partial \theta }},
\end{align}
and the other 
\begin{align}
&\left( {V_1^e } \right)_{nlm}  = a\frac{{\sqrt {l\left( {l + 1} \right)} }}{n}\frac{{\Pi _{nl} \left( \chi  \right)}}{{\sin \chi }}Y_{lm} \left( {\theta ,\varphi } \right), \nonumber\\&
 \left( {V_2^e } \right)_{nlm}  = \frac{a}{{n\sqrt {l\left( {l + 1} \right)} }}\bigg[\left( {l + 1} \right)\cos \chi \Pi _{nl} \left( \chi  \right)-\sqrt {n^2  - \left( {l + 1} \right)^2 } \sin \chi \Pi _{nl + 1} \left( \chi  \right) \bigg]\frac{{\partial Y_{lm} }}{{\partial \theta }},  \nonumber\\&
 \left( {V_3^e } \right)_{nlm}  = \frac{a}{{n\sqrt {l\left( {l + 1} \right)} }}\bigg[ \left( {l + 1} \right)\cos \chi \Pi _{nl} \left( \chi  \right) - \sqrt {n^2  - \left( {l + 1} \right)^2 } \sin \chi \Pi _{nl + 1} \left( \chi  \right) \bigg]\frac{{\partial Y_{lm} }}{{\partial \varphi }}.
\end{align}
One can show that
\begin{align}
&\int\limits_{S^3 \left( a \right)} d\mu \bar g^{ij} \left( {V_i^o } \right)_{nlm} \left( {V_j^o } \right)^* _{n'l'm'} =\nonumber \\& 
\int\limits_{S^3 \left( a \right)} d\mu \bar g^{ij} \left( {V_i^e } \right)_{nlm} \left( {V_j^e } \right)^* _{n'l'm'}  = \delta _{nn'} \delta _{ll'} \delta _{mm'},
 \end{align}
 and
 \begin{equation}
\bar g^{ij} \left( {V_i^o } \right)_{nlm} \left( {V_j^e } \right)_{n'l'm'}  = 0 .
 \end{equation}
These vector harmonics constitute a complete orthonormal set for the expansion of any transverse vector field on $S^3(a)$. Thus, for vector perturbation $A_i(\textbf{x})$ we can write
 \begin{equation}
A_i \left( {\bf{x}} \right) = \sum\limits_{nlm} {\Big[  {A_{nlm}^o \left( {V_i^o } \right)_{nlm}  + A_{nlm}^e \left( {V_i^e } \right)_{nlm} }\Big]  } ,
 \end{equation}
where $A^o_{nlm} $ and $ A^e_{nlm} $ are two random fields and like scalar perturbations we have
\begin{align}
& \langle A_{nlm}^o A_{n'l'm'}^{o*}  \rangle  = P_A^o \left( n \right)\delta _{nn'} \delta _{ll'} \delta _{mm'}, \nonumber\\ \nonumber\\&
\langle A_{nlm}^e A_{n'l'm'}^{e *}  \rangle  = P_A^e \left( n \right)\delta _{nn'} \delta _{ll'} \delta _{mm'},
   \nonumber\\ \nonumber\\& 
\langle A_{nlm}^o A_{n'l'm'}^{e *}  \rangle  = P_A^{oe} \left( n \right)\delta _{nn'} \delta _{ll'} \delta _{mm'} . \nonumber
\end{align}
It yields
\begin{align}
& \langle A_i(\textbf{x})A_j^*(\textbf{x})\rangle = \nonumber \\&
 \quad\sum\limits_{nlm} \Big[P_A^o(n)(V_i^o(\textbf{x}))_{nlm}  (V_i^{o*}(\textbf{x}))_{nlm} +P_A^e(n)(V_i^e(\textbf{x}))_{nlm} (V_i^{e*}(\textbf{x}))_{nlm}\nonumber \\&
 \quad +P_A^{oe}(n)(V_i^o(\textbf{x}))_{nlm}(V_i^{e*}(\textbf{x}))_{nlm} +P_A^{oe}(n)(V_i^e(\textbf{x}))_{nlm} 
 (V_i^{o*}(\textbf{x}))_{nlm}\Big].
\end{align}
On the other hand, $ \langle A_i(\textbf{x})A_j^*(\textbf{x})\rangle $ must not change under parity transformation, because probability distribution function is invariant under spatial inversion so, $ P^{oe}_A(n)=0 $. Furthermore,
\begin{equation}
P_A^o(n)=P_A^e(n)=P_A^{+1}(n) ,
\end{equation}
Because the power spectrum just depends on the probability distribution function and it cannot be function of parity. Thus,
\begin{align}
&\langle A^o_{nlm} A^{o*}_{nlm}\rangle =\langle A^e_{nlm} A^{e*}_{nlm}\rangle =P_A^{+1}(n)\delta_{nn'}\delta_{ll'}\delta_{mm'}\\ \nonumber \\&
\langle A^o_{nlm} A^{e*}_{nlm}\rangle=0 .
\end{align}
The last relation means $A^o_{nlm} $ and $A^e_{nlm} $ are statistically uncorrelated random fields, however, they have the same spectrum.
\subsection{Tensor perturbations and tensor random fields}
Every symmetric, traceless and transverse covariant tensor of rank 2 on $ S^3(a) $ can be expanded in terms of $ t-t $ tensor spherical harmonics \cite{Ref22}. These harmonics can be classified into two groups:\\
\textit{Odd parity} 
\begin{align}
\begin{split}
 \left( {T_{11}^o } \right)_{nlm}  = &0, \\ \\
 \left( {T_{22}^o } \right)_{nlm}  =  &- \frac{{a^2 }}{{\sqrt {2\left( {n^2  - 1} \right)l\left( {l - 1} \right)\left( {l + 1} \right)\left( {l + 2} \right)} }}\times\\ \sin \chi & \bigg[ {\left( {l + 2} \right)\cos \chi \Pi _{nl} \left( \chi  \right) - \sqrt {n^2  - \left( {l + 1} \right)^2 } \sin \chi \Pi _{nl + 1} \left( \chi  \right)} \bigg] \frac{{X_{lm} \left( {\theta ,\varphi } \right)}}{{\sin \theta }}, \\ \\
 \left( {T_{33}^o } \right)_{nlm}  =&\frac{{a^2 }}{{\sqrt {2\left( {n^2  - 1} \right)l\left( {l - 1} \right)\left( {l + 1} \right)\left( {l + 2} \right)} }}\times\\ \sin \chi &\bigg [ {\left( {l + 2} \right)\cos \chi \Pi _{nl} \left( \chi  \right) - \sqrt {n^2  - \left( {l + 1} \right)^2 } \sin \chi \Pi _{nl + 1} \left( \chi  \right)}\bigg]\sin \theta X_{lm} \left( {\theta ,\varphi } \right), \\ \\
 \left( {T_{12}^o } \right)_{nlm}  =& - a^2 \sqrt {\frac{{\left( {l - 1} \right)\left( {l + 2} \right)}}{{2\left( {n^2  - 1} \right)l\left( {l + 1} \right)}}} \Pi _{nl} \left( \chi  \right) \frac{1}{{\sin \theta }}\frac{{\partial Y_{lm} }}{{\partial \varphi }}, \\ \\
 \left( {T_{13}^o } \right)_{nlm}  = & a^2 \sqrt {\frac{{\left( {l - 1} \right)\left( {l + 2} \right)}}{{2\left( {n^2  - 1} \right)l\left( {l + 1} \right)}}}\Pi _{nl} \left( \chi  \right)\sin \theta \frac{{\partial Y_{lm} }}{{\partial \theta }}, \\ \\ 
 \left( {T_{23}^o } \right)_{nlm}  = & \frac{{a^2 }}{{\sqrt {2\left( {n^2  - 1} \right)l\left( {l - 1} \right)\left( {l + 1} \right)\left( {l + 2} \right)} }}\times\\&\sin \chi\bigg[ {\left( {l + 2} \right)\cos \chi \Pi _{nl} \left( \chi  \right)-  \sqrt {n^2  - \left( {l + 1} \right)^2 } \sin \chi \Pi _{nl + 1} \left( \chi  \right)}\bigg]\sin \theta W_{lm} \left( {\theta ,\varphi } \right),
 \end{split}
 \end{align} 
 where
\begin{align}
 &X_{lm} \left( {\theta ,\varphi } \right) = 2\left( {\frac{{\partial ^2 Y_{lm} }}{{\partial \theta \partial \varphi }} - \cot \theta \frac{{\partial Y_{lm} }}{{\partial \varphi }}} \right), \\& 
 W_{lm} \left( {\theta ,\varphi } \right) = 2\frac{{\partial ^2 Y_{lm} }}{{\partial \theta ^2 }} + l\left( {l + 1} \right)Y_{lm} \left( {\theta ,\varphi } \right) .
 \end{align}
 \textit{Even parity}
\begin{align}
\begin{split}
 \left( {T_{11}^e } \right)_{nlm}  = &\frac{{a^2 }}{n}\sqrt {\frac{{l\left( {l - 1} \right)\left( {l + 1} \right)\left( {l + 2} \right)}}{{2\left( {n^2  - 1} \right)}}} \frac{{\Pi _{nl} \left( \chi  \right)}}{{\sin ^2 \chi }}Y_{lm} \left( {\theta ,\varphi } \right), \\ \\
 \left( {T_{22}^e } \right)_{nlm}  =  &- \frac{{a^2 }}{{2n}}\sqrt {\frac{{l\left( {l - 1} \right)\left( {l + 1} \right)\left( {l + 2} \right)}}{{2\left( {n^2  - 1} \right)}}} \Pi _{nl} \left( \chi  \right)Y_{lm} \left( {\theta ,\varphi } \right) \\ &+ \frac{{a^2 }}{{n\sqrt {2\left( {n^2  - 1} \right)l\left( {l - 1} \right)\left( {l + 1} \right)\left( {l + 2} \right)} }}G_{nl} \left( \chi  \right) W_{lm} \left( {\theta ,\varphi } \right),\\ \\
 \left( {T_{33}^e } \right)_{nlm}  = &- \frac{{a^2 }}{{2n}}\sqrt {\frac{{l\left( {l - 1} \right)\left( {l + 1} \right)\left( {l + 2} \right)}}{{2\left( {n^2  - 1} \right)}}} \Pi _{nl} \left( \chi  \right)\sin ^2 \theta Y_{lm} \left( {\theta ,\varphi } \right)- \\&\frac{{a^2 }}{{n\sqrt {2\left( {n^2  - 1} \right)l\left( {l - 1} \right)\left( {l + 1} \right)\left( {l + 2} \right)} }} G_{nl} \left( \chi  \right)\sin ^2 \theta W_{lm} \left( {\theta ,\varphi } \right), \\ \\
 \left( {T_{12}^e } \right)_{nlm}  =&\frac{{a^2 }}{n}\sqrt {\frac{{\left( {l - 1} \right)\left( {l + 2} \right)}}{{2\left( {n^2  - 1} \right)l\left( {l + 1} \right)}}} \times\\& \bigg[ {\left( {l + 1} \right)\cot \chi \Pi _{nl} \left( \chi  \right)- \sqrt {n^2  - \left( {l + 1} \right)^2 } \Pi _{nl + 1} \left( \chi  \right)} \bigg]\frac{{\partial Y_{lm} }}{{\partial \theta }}, \\ \\
 \left( {T_{13}^e } \right)_{nlm}  =& \frac{{a^2 }}{n}\sqrt {\frac{{\left( {l - 1} \right)\left( {l + 2} \right)}}{{2\left( {n^2  - 1} \right)l\left( {l + 1} \right)}}}\times\\&\bigg[ {\left( {l + 1} \right)\cot \chi \Pi _{nl} \left( \chi  \right)- \sqrt {n^2  - \left( {l + 1} \right)^2 } \Pi _{nl + 1} \left( \chi  \right)} \bigg] \frac{{\partial Y_{lm} }}{{\partial \varphi }}, \\ \\
 \left( {T_{23}^e } \right)_{nlm}  = & \frac{{a^2 }}{{n\sqrt {2\left( {n^2  - 1} \right)l\left( {l - 1} \right)\left( {l + 1} \right)\left( {l + 2} \right)} }}G_{nl} \left( \chi  \right)X_{lm} \left( {\theta ,\varphi } \right), \\ 
\end{split} 
\end{align}
where
\begin{align}
G_{nl} \left( \chi  \right) =& \left( {l + 2} \right)\cos ^2 \chi \Pi _{nl} \left( \chi  \right) - \left( {n^2  - 1} \right)\sin ^2 \chi \Pi _{nl} \left( \chi  \right)&\nonumber\\ &
+ \frac{1}{2}\left( {l - 1} \right)\left( {l + 2} \right)\Pi _{nl} \left( \chi  \right) - \sqrt {n^2  - \left( {l + 1} \right)^2 } \sin \chi \cos \chi \Pi _{nl + 1} \left( \chi  \right).
\end{align}
It is also possible to express the tensor harmonics on $ S^3(a) $ in terms of the \textit{Chebyshev polynomials of the first kind} \cite{Ref31} which constitute the tensor Cartesian harmonics.
It can be shown 
\begin{align}
& \nabla ^2 \left( {T_{ij}^o } \right)_{nlm}  = \frac{{3 - n^2 }}{{a^2 }}\left( {T_{ij}^o } \right)_{nlm}, \quad n = 3,4,... \\ &
 \nabla ^2 \left( {T_{ij}^e } \right)_{nlm}  = \frac{{3 - n^2 }}{{a^2 }}\left( {T_{ij}^e } \right)_{nlm},\quad n = 3,4,... 
 \end{align}
and also
\begin{align}
 &\int\limits_{S^3 \left( a \right)} d\mu \bar g^{ik} \bar g^{jl} \left( {T_{ij}^o } \right)_{nlm} \left( {T_{kl}^o } \right)^* _{n'l'm'}= \nonumber\\&
  \int\limits_{S^3 \left( a \right)} d\mu \bar g^{ik} \bar g^{jl} \left( {T_{ij}^e } \right)_{nlm} \left( {T_{kl}^e } \right)^* _{n'l'm'} = \delta _{nn'} \delta _{ll'} \delta _{mm'} . 
 \end{align}
The set $\lbrace(T_{ij}^o)_{nlm}, (T_{ij}^e)_{nlm}\rbrace$ constitutes a complete orthonormal basis for the expansion of any symmetric traceless-divergence-free covariant tensor field of rank 2 on $ S^3(a) $. On the other hand, the tensor mode is completely characterized by two traceless-transverse symmetric tensors $D_{ij}(t,\textbf{x}) $ and $ \Pi^T_{ij}(t,\textbf{x})$. We can expand them in terms of $ t-t $ tensor spherical harmonics on $ S^3(a) $ :
\begin{equation}
D_{ij} \left( {\bf{x}} \right) = \sum\limits_{nlm} {\left[ {D_{nlm}^o \left( {T_{ij}^o } \right)_{nlm}  + D_{nlm}^e \left( {T_{ij}^e } \right)_{nlm} } \right]} .
\end{equation}
There is a similar expansion for $ \Pi^T_{ij}(t,\textbf{x}) $ (Note that we drop $ t $ here, because all quantities are considered at a fixed instant). ${D_{nlm}^o } $ and $ {D_{nlm}^e } $ just like  $D_{ij}(t,\textbf{x}) $ are two random fields, so
\begin{align}
\langle D_{nlm}^o &D_{n'l'm'}^{o*}  \rangle  =  \langle D_{nlm}^e D_{n'l'm'}^{e*}  \rangle = P_D^{ + 2} \left( n \right)\delta _{nn'} \delta _{ll'} \delta _{mm'} ,
\end{align}
where $ P_D^{ + 2} \left( n \right) $ is the power spectrum of the gravitational wave $ D_{ij} $ \cite{Ref32}. The probability distribution is independent of parity, so we cannot expect $ \langle D_{nlm}^o D_{n'l'm'}^{o*}  \rangle $ and $ \langle D_{nlm}^e D_{n'l'm'}^{e*}  \rangle $ having different values. In addition, because scalar, vector and tensor modes are independent, their joint power spectrums are vanished.

\section{The gauge problem}
\label{sec4}
In this section, we investigate the behavior of the perturbations under the gauge transformations. The equations derived in Section \ref{sec2} may have physically equivalent solutions. This problem is called gauge freedom. Similar to the Einstein's field equations this gauge freedom may be fixed by choosing a coordinate system. For this purpose, let's consider a spacetime coordinate transformation
\begin{equation}\label{a1}
x^\mu   \to x'^\mu   = x^\mu   + \epsilon ^\mu  \left( x \right),
\end{equation}
with small $ \epsilon ^\mu  \left( x \right) $ in the same sence that $ h_{\mu\nu} $ and other perturbations are small. In cosmology, we call Eq.(\ref{a1}) a gauge transformation, if it affects only the field perturbations and preserves unperturbed metric\cite{Ref2,Ref33}. Under such gauge transformation, the metric of spacetime changes as
\begin{equation}
g_{\mu \nu } \left( x \right) \to g'_{\mu \nu } \left( {x'} \right) = \frac{{\partial x^\rho  }}{{\partial x'^\mu  }}\frac{{\partial x^\lambda  }}{{\partial x'^\nu  }}g_{\rho \lambda } \left( x \right),
\end{equation}
equivalently
\begin{equation}
g_{\mu \nu } \left( x \right) = \frac{{\partial x'^\rho  }}{{\partial x^\mu  }}\frac{{\partial x'^\lambda  }}{{\partial x^\nu  }}g'_{\rho \lambda } \left( {x + \epsilon } \right).
\end{equation}
It yields 
\begin{align}
\bar g_{\mu \nu } \left( x \right) + h_{\mu \nu } \left( x \right) &= \left( {\delta ^\rho \smallskip_\mu   + \partial _\mu  \epsilon ^\rho  } \right)\left( {\delta^\lambda \smallskip _\nu    + \partial _\nu  \epsilon ^\lambda  } \right)\left[ {\bar g_{\rho \lambda } \left( {x + \epsilon } \right) + h'_{\rho \lambda } \left( x \right)} \right].
\end{align}
After simplification we have
\begin{equation}
h'_{\mu \nu } \left( x \right) = h_{\mu \nu } \left( x \right) - \epsilon ^\lambda  \left( {\partial _\lambda  \bar g_{\mu \nu } } \right) - \bar g_{\mu \lambda } \left( {\partial _\nu  \epsilon ^\lambda  } \right) - \bar g_{\nu \lambda } \left( {\partial _\mu  \epsilon ^\lambda  } \right).
\end{equation}
Thus
\begin{equation}
\Delta h_{\mu \nu } \left( x \right) = h'_{\mu \nu } \left( x \right) - h_{\mu \nu } \left( x \right) =  - \boldsymbol{\nabla} _\mu  \epsilon _\nu   - \boldsymbol{\nabla} _\nu  \epsilon _\mu ,
\end{equation}
where $ \boldsymbol{\nabla}_\mu $ is the covariant derivative corresponding to $ \bar g_{\mu \nu }$. Consequently
\begin{align}
 &\Delta h_{00}  =  - 2\dot \epsilon _0 \label{a2} ,\\ &
 \Delta h_{i0}  = \Delta h_{0i}  =  - \dot \epsilon _i  - \partial _i \epsilon _0  + 2\frac{{\dot a}}{a}\epsilon _i \label{a3}, \\ &
 \Delta h_{ij}  =  - \nabla _i \epsilon _j  - \nabla _j \epsilon _i  + 2a\dot a\tilde g_{ij} \epsilon _0 ,  \label{a4}
 \end{align}
where $ \nabla_i $ is the covariant derivative respect to $ \bar g_{ij}$.\\
Similarly we can derive the effect of gauge transformation Eq.(\ref{a1}) on the energy-momentum tensor
\begin{equation}
\Delta \left( {\delta T_{\mu \nu } } \right) =  - \epsilon ^\lambda  \left( {\partial _\lambda  \bar T_{\mu \nu } } \right) - \bar T_{\mu \lambda } \left( {\partial _\nu  \epsilon ^\lambda  } \right) - \bar T_{\nu \lambda } \left( {\partial _\mu  \epsilon ^\lambda  } \right) ,
\end{equation}
or in more detail
\begin{align}
& \Delta \left( {\delta T_{00} } \right) = 2\bar \rho \dot \epsilon _0  +\mathop {\bar \rho}\limits^ .  \epsilon _0,  \label{a5}\\ &
 \Delta \left( {\delta T_{i0} } \right) = \Delta \left( {\delta T_{0i} } \right) = 2\bar p\frac{{\dot a}}{a}\epsilon _i  - \bar p\dot \epsilon _i  + \bar \rho\partial_i \epsilon _0 \label{a6} , \\ &
 \Delta \left( {\delta T_{ij} } \right) =  - \bar p\left( {\nabla _i \epsilon _j  + \nabla _j \epsilon _i } \right) + \frac{d}{{dt}}\left( {a^2 \bar p} \right)\tilde g_{ij} \epsilon _0 . \label{a7}
 \end{align}
 In order to derive the gauge transformations of the scalar, vector and tensor parts of $h_{\mu \nu} $ and $T_{\mu \nu} $, it is necessary to decompose the spatial part of $ \epsilon^\mu $ as follows
\begin{equation}
\epsilon _i  = \nabla _i \epsilon ^S  +\epsilon_i^V ,\quad \nabla ^i \epsilon _i^V  = 0. \label{a8}
\end{equation}
Now  with substitution Eq.(\ref{a8}) in Eqs.(\ref{a2}), (\ref{a3}), (\ref{a4}), (\ref{a5}), (\ref{a6}) and (\ref{a7}), we find
\begin{align}
\begin{array}{l}
 \Delta A = 2\frac{{\dot a}}{a}\epsilon _0 ,\quad \Delta B =  - \frac{2}{{a^2 }}\epsilon ^S , \\ \\
 \Delta E = 2\dot \epsilon _0 , \quad \Delta F = \frac{1}{a}\left( { - \dot \epsilon ^S  - \epsilon _0  + 2\frac{{\dot a}}{a}\epsilon ^S } \right), \\  \\
 \Delta C_i  =  - \frac{1}{{a^2 }}\epsilon _i^V ,\quad \Delta G_i  = \frac{1}{a}\left( { - \dot \epsilon _i^V  + 2\frac{{\dot a}}{a}\epsilon _i^V } \right), \\  \\
 \Delta D_{ij}  = 0 , \quad \Delta \Pi ^S  = \Delta \Pi _i^V  = \Delta \Pi _{ij}^T  = 0, \\ \\
 \Delta \delta u =  - \epsilon _0 ,\quad \Delta \delta u_i^V  = 0, \\ \\
 \Delta \delta \rho  = \mathop {\bar \rho}\limits^ .  \epsilon _0 , \quad \Delta \delta p = \mathop {\bar p}\limits^ . \epsilon _0.  \\  \\
 \end{array}
\end{align}
Obviously $ \Pi^S , \Pi^V_i , \Pi^T_{ij}, D_{ij} $ and $\delta u^V_i  $ are gauge invariant quantities. Besides, one can construct more gauge invariant quantities by combination of the perturbative quantities, e.g. $ \zeta  = \frac{A}{2} - H\frac{{\delta \rho }}{{\mathop {\bar \rho}\limits^ . }}$ $ (H = \frac{{\dot a}}{a} )$ which is known as the\textit{curvature perturbation on the uniform density slices} \cite{Ref34,Ref35}. Note in particular that $ \zeta $ is a pivotal quantity in cosmology which is related to the fluctuations of inflaton as well as, CMB angular power spectrum \cite{Ref32,Ref36} and consequently connects the primordial perturbations to the present observational data. \\
All of the tensor quantities are gauge invariant and in result gauge-fixing is not required. On the other hand, for the vector mode, we can fix a gauge by choosing $ \epsilon^V_i $ so that either $C_i $ or $ G_i$ vanishes. For the scalar perturbations, fixing a gauge means choosing $\epsilon_0 $ and $ \epsilon^S $, so there are several ways to fix a gauge \cite{Ref5}, but here we concentrate on a special gauge which was introduced by Mukhanov et al. \cite{Ref37} and is known as \textit{Newtonian gauge}. In this gauge we choose  $\epsilon_0 $ and $ \epsilon^S $ by setting $B=F=0$. It is convenient to write $E$ and $A$ in this gauge as
\begin{equation}
E = 2\Phi ,\quad A =  - 2\Psi .
\end{equation}
$\Phi $ and  $\Psi $ are known as \textit{Bardeen's potentials} \cite{Ref34}. This gauge eliminates the gauge freedom completely in contrast to the synchronous gauge \cite{Ref2,Ref38} which was introduced first time by Lifshitz \cite{Ref3}. In the Newtonian gauge the line element of the universe takes the form
\begin{equation}
ds^2  =  - \left( {1 + 2\Phi } \right)dt^2  + a^2 \tilde g_{ij} \left( {1 - 2\Psi } \right)dx^i dx^j ,
\end{equation}
and the gravitational field and conservation equations become
\begin{align}
&  - \frac{4}{{a^2 }}\Psi  + 6H\dot \Psi  + \ddot \Psi  + 2\left( {3H^2  + \dot H} \right)\Phi  + H\dot \Phi  - \nabla ^2 \Psi \nonumber \\&
\quad = 4\pi G\left( { - \delta \rho  + \delta p + a^2 \nabla ^2 \Pi ^S } \right), \\ &
 \Psi  - \Phi  = 8\pi G a^2 \Pi ^S, \label{a9} \\ &
 \dot \Psi  + H\Phi  =  - 4\pi G\left( {\bar \rho  + \bar p} \right)\delta u , \\ &
 3\ddot \Psi  + 6H\dot \Psi  + 3H\dot \Phi  + \nabla ^2 \Phi  + 6\left( {H^2  + \dot H} \right)\Phi \nonumber\\&
  \quad= 4\pi G\left( {\delta \rho  + 3\delta p + a^2 \nabla ^2 \Pi ^S } \right) ,\\ &
 3\left( {\bar \rho  + \bar p} \right)\dot \Psi  = \frac{{\partial \delta \rho }}{{\partial t}} + 3H\left( {\delta \rho  + \delta p} \right) +\nabla ^2 \left[ {\left( {\bar \rho  + \bar p} \right)\delta u + a^2 H\Pi ^S } \right] \\ &
 \left( {\bar \rho  + \bar p} \right)\Phi  =  - \mathop {\bar p}\limits^ . \delta u - \left( {\bar \rho  + \bar p} \right)\frac{{\partial \delta u}}{{\partial t}} - \delta p - a^2 \nabla ^2 \Pi ^S- 2\Pi ^S.  
 \end{align}
 In the next section we shall show that this system of equations has two independent adiabatic solutions.
 
 \section{Adiabatic modes in a spatially closed universe}
\label{sec5}
In this section, we want to generalize the Weinberg's theorem\cite{Ref2,Ref39} which has been proved for a spatially flat universe to the spatially closed case. According to this theorem whatever the contents of the universe, the perturbative field equations have two independent \textit{adiabatic} solutions in the time intervals when the perturbation scales are often very longer than the Hubble horizon of the universe. These two solutions in Newtonian gauge are
\begin{equation}
\left\{ \begin{split}
 &\Psi \left( {t,{\bf{x}}} \right) = \Phi \left( {t,{\bf{x}}} \right) = \zeta \left( {\bf{x}} \right)\left[ {\frac{H}{a}\int\limits_{t_0 }^t {a\left( \tau  \right)d\tau  - 1} } \right] , \\& 
 \frac{{\delta \rho \left( {t,{\bf{x}}} \right)}}{{\mathop {\bar \rho}\limits^ .  }} = \frac{{\delta p\left( {t,{\bf{x}}} \right)}}{{\mathop {\bar p}\limits^ . }} =  - \delta u\left( {t,{\bf{x}}} \right) =  - \frac{{\zeta \left( {\bf{x}} \right)}}{a}\int\limits_{t_0 }^t {a\left( \tau  \right)d\tau } ,  \\ &
 \Pi ^S \left( {t,{\bf{x}}} \right) = 0 , \\ 
\end{split} \right.\\ 
\end{equation}
and
\begin{equation}
 \left\{ \begin{split}
 &\Psi \left( {t,{\bf{x}}} \right) = \Phi \left( {t,{\bf{x}}} \right) =  \chi \left( {\bf{x}} \right)\frac{H}{a} ,  \\& 
 \frac{{\delta \rho \left( {t,{\bf{x}}} \right)}}{{\mathop {\bar \rho}\limits^ .  }} = \frac{{\delta p\left( {t,{\bf{x}}} \right)}}{{\mathop {\bar p}\limits^ . }} =  - \delta u\left( {t,{\bf{x}}} \right) =  - \frac{{\chi \left( {\bf{x}} \right)}}{a} , \\ &
 \Pi ^S \left( {t,{\bf{x}}} \right) = 0 , \\ 
 \end{split} \right.\\
\end{equation}
which $ \zeta(\textbf{x}) $ is the curvature perturbation on the uniform density slices when the perturbations are outside of the Hubble horizon or equivalently \textit{conformal factor} of $ S^3 $ and $ \chi(\textbf{x}) $ is an arbitrary function of position.\\ 
In order to prove, initially we put $ \Pi^S=0 $, because the cosmic fluid is approximately perfect; thus, from Eq.(\ref{a9}) we have
\begin{equation}\label{b1}
\Psi=\Phi ,
\end{equation}
Now suppose the gauge transformation
\begin{equation}
x^\mu   \to x^\mu   + \epsilon ^\mu  \left( x \right) ,
\end{equation}
which converts the present Newtonian gauge to another Newtonian gauge. Consequently
\begin{align}
& \Delta h_{00}  =  - 2\dot \epsilon _0  \Rightarrow \Delta \Phi  = \dot \epsilon _0 , \label{b2}\\ &
 \Delta h_{i0}  = 0 \Rightarrow  - \dot \epsilon _i  - \partial _i \epsilon _0  + 2\frac{{\dot a}}{a}\epsilon _i  = 0 , \label{b3}\\ &
 \Delta h_{ij}  =  - 2a^2 \tilde g_{ij} \Delta \Psi  \Rightarrow  - \nabla _i \epsilon _j  - \nabla _j \epsilon _i  + 2a\dot a\tilde g_{ij} \epsilon _0   \nonumber\\&\qquad=  - 2a^2 \tilde g_{ij} \Delta \Psi .\label{b4} 
 \end{align}
Eq.(\ref{b3}) results in
\begin{equation}\label{b5}
\epsilon _i \left( {t,{\bf{x}}} \right) =  - a^2 \int\limits_{t_0 }^t {\frac{{\partial _i \epsilon _0 \left( {\tau ,{\bf{x}}} \right)}}{{a^2 \left( \tau  \right)}}} d\tau  + a^2 \eta _i \left( {\bf{x}} \right) ,
\end{equation}
which $t_0	$ and $\eta_i(\textbf{x}) $ respectively are arbitrary time and arbitrary 3-vector field on $ S^3 $. Substituting Eq.(\ref{b5}) in Eq.(\ref{b4}) yields
\begin{align}\label{b6}
&2 \int\limits_{t_0 }^t {\frac{{H _{ij} \epsilon _0 \left( {\tau ,{\bf{x}}} \right)}}{{a^2 \left( \tau  \right)}}} d\tau  -  \left( {\nabla _i \eta _j  + \nabla _j \eta _i } \right) \nonumber\\ & + 2H\tilde g_{ij} \epsilon _0 \left( {t,{\bf{x}}} \right)=  - 2 \tilde g_{ij} \Delta \Psi .
\end{align}
Now suppose that $\eta_i(\textbf{x}) $ is a \textit{conformal Killing vector} of $ S^3 $
\begin{equation}\label{b7}
\nabla _i \eta _j  + \nabla _j \eta _i  = 2\gamma \left( {\bf{x}} \right)\tilde g_{ij} ,
\end{equation}
where $ \gamma \left( {\bf{x}} \right)=\frac{1}{3}\nabla_i\eta^i $ is a function on $ S^3 $ so-called \textit{conformal factor } of $ S^3 $ \cite{Ref40}. Note that $ S^3 $ has no any\textit{ homothetic Killing vector} \cite{Ref40,Ref41}, but due to its conformal symmetry, it has conformal Killing vector. Indeed in \cite{Ref42} has been proved that  $ S^3 $ has four gradient  conformal Killing vector. For instance, $\eta _i  = \delta ^m\smallskip_ i  \smallskip (m = 1,2,3) $ is a conformal Killing vector of $ S^3 $ with conformal factor $ -x^m $:
\begin{equation*}
\nabla _i \delta ^m\smallskip _ j  + \nabla _j \delta ^m \smallskip _ i =  - 2x^m \tilde g_{ij} .
\end{equation*}
On the other hand, in the super-Hubble scales we can ignore the first term on the left side of Eq.(\ref{b6}), because $ 2 \int\limits_{t_0 }^t {\frac{{H_{ij} \epsilon _0 \left( {\tau ,{\bf{x}}} \right)}}{{a^2 \left( \tau  \right)}}} d\tau $ is of the order of $ 2 \int\limits_{t_0 }^t {\nabla^2 \epsilon _0 \left(\tau ,\bf{x}\right) d\tau} $, so its Fourier transform has same order of $2\int\limits_{t_0 }^t {\frac{{1 - n^2 }}{{a^2 \left( \tau  \right)}} \epsilon_{0_{nlm}} \left( \tau \right)d\tau} $ which is negligible for super-Hubble scales. Thus Eq.(\ref{b6}) in the time intervals when the perturbation scales are very longer than the Hubble horizon, turns to 
\begin{equation*}
 - \left( {\nabla _i \eta _j  + \nabla _j \eta _i } \right) + 2H\tilde g_{ij} \epsilon _0 \left( {t,{\bf{x}}} \right) =  - 2 \tilde g_{ij} \Delta \Psi ,
\end{equation*}
or
\begin{equation}\label{b8}
\Delta \Psi  = \gamma \left( {\bf{x}} \right) - H\epsilon _0 \left( {t,{\bf{x}}} \right).
\end{equation}
Besides, in the Newtonian gauge both $\Psi $ and $\Psi+\Delta\Psi $ are solutions, so that it results from the linearity of equations, $ \Delta\Psi $ is another solution of the Newtonian field equations too. It is also true for other perturbations. Consequently, we have a set of solutions of the Newtonian gauge field equations:
\begin{align}
 &\Psi  = \gamma \left( {\bf{x}} \right) - H\epsilon _0 \left( {t,{\bf{x}}} \right) ,\label{b9} \\& 
 \Phi  = \dot \epsilon _0 \left( {t,{\bf{x}}} \right) ,\label{c1} \\ &
 \delta \rho  = \mathop {\bar \rho}\limits^ .  \epsilon _0 \left( {t,{\bf{x}}} \right) ,\label{c2} \\& 
 \delta p = \mathop {\bar p}\limits^ . \epsilon _0 \left( {t,{\bf{x}}} \right) ,\label{c3} \\& 
 \delta u =  - \epsilon _0 \left( {t,{\bf{x}}} \right) .\label{c4} 
 \end{align}
Furthermore,
\begin{equation}\label{c5}
\zeta  =  - \Psi  - H\frac{{\delta \rho }}{{\mathop {\bar \rho}\limits^ .  }} =  - \gamma \left( {\bf{x}} \right).
\end{equation}
It can be concluded from Eq.(\ref{c5}) that $ \zeta $ is conserved i.e. it doesn't depend on the time, so that above solutions are appropriate to a period when the perturbations are outside of the Hubble horizon. In order to see conservation of $ \zeta $ in the super-Hubbles scales, it is sufficient to write the Fourier transformation of Eq.(\ref{g})
\begin{align}
\begin{split}
&\frac{{\partial \delta \rho _n }}{{\partial t}} + \frac{{1 - n^2 }}{{a^2 }}\Big[ { - a\left( {\bar \rho  + \bar p} \right)F_n  + \left( {\bar \rho  + \bar p} \right)\delta u_n  + a\dot a\Pi _n^S } \Big] \\ &
 + \frac{3}{2}\left( {\bar \rho  + \bar p} \right)\dot A_n  + \frac{1}{2}\left( {\bar \rho  + \bar p} \right)\left( {1 - n^2 } \right)\dot B_n + 3\frac{{\dot a}}{a}\left( {\delta \rho _n  + \delta p_n } \right) = 0 .
 \end{split}
\end{align}
for simplicity we drop $l $ and $m $ indices. On the super-Hubble scales $ (n<<aH) $ we can approximate this equation as follows
\begin{equation}\label{d1}
\frac{{\partial \delta \rho _n }}{{\partial t}} + \frac{3}{2}\left( {\bar \rho  + \bar p} \right)\dot A_n  + 3\frac{{\dot a}}{a}\left( {\delta \rho _n  + \delta p_n } \right) = 0 .
\end{equation}
On the other hand, we have
\begin{equation}\label{d2}
A_n  = 2\zeta _n  - \frac{2}{3}\frac{{\delta \rho _n }}{{\bar \rho  + \bar p}},
\end{equation}
By substituting Eq.(\ref{d2}) in Eq.(\ref{d1}) and using conservation law of energy in unperturbed universe we can write
\begin{equation}
\dot \zeta _n  = \frac{{\mathop {\bar p}\limits^ . \delta \rho _n  - \mathop {\bar \rho}\limits^ .  \delta p_n }}{{3\left( {\bar \rho  + \bar p} \right)^2 }} .
\end{equation}
Thus for adiabatic perturbations for which $ \frac{{\delta \rho _n }}{{\mathop {\bar \rho}\limits^ .}} = \frac{{\delta p_n }}{{\mathop{\bar p}\limits^ .}} $, we have
\begin{equation}
\dot \zeta _n  = 0 .
\end{equation}
Consequently, if the perturbations are adiabatic\footnote{Strictly speaking, the adiabatic condition is $\frac{\delta {\rho _\alpha }}{{\mathop{\bar\rho}\limits^. }_\alpha} = \frac{\delta {\rho _\beta}}{{\mathop{\bar\rho}\limits^. }_\beta}$ where $ \alpha $ and $ \beta $ stand for every two different species of cosmic fluid elements whereas the condition $ \frac{{\delta \rho _n }}{{\mathop {\bar \rho}\limits^ .}} = \frac{{\delta p_n }}{{\mathop{\bar p}\limits^ .}} $ is known as the \textit{generalized} adiabatic condition.}, $ \zeta $ is conserved of course in the epoch when the wavelength of most perturbations are very longer than the Hubble radius. Indeed, the conservation of $ \zeta $ is a general theorem in cosmology which has been proved even for nonlinear generalization of $ \zeta $ \cite{Ref43}. Note that ignoring the first term of the left hand side of Eq.(\ref{b6}) causes $ \zeta $ to be independent of time which is equivalent to going outside of the Hubble horizon.\\
From combination of Eqs.(\ref{b1}), (\ref{b9}), (\ref{c1}) and also Eq.(\ref{c5}) we may write
\begin{equation}\label{c7}
\dot \epsilon _0 \left( {t,{\bf{x}}} \right) + H\epsilon _0 \left( {t,{\bf{x}}} \right) =  - \zeta \left( {\bf{x}} \right).
\end{equation}
Eq.(\ref{c7}) is a first order differential equation for $\epsilon _0 \left( {t,{\bf{x}}} \right) $ and we solve it in two different cases:
At first we assume $ \zeta \left( {\bf{x}} \right) \ne 0 $ consequently, Eq.(\ref{c7}) results in
\begin{equation}\label{c8}
\epsilon _0 \left( {t,{\bf{x}}} \right) =  - \frac{{\zeta \left( {\bf{x}} \right)}}{a}\int\limits_{t_0 }^t {a\left( \tau  \right)d\tau } .
\end{equation}
By inserting Eq.(\ref{c8}) in Eqs.(\ref{b9})-(\ref{c4}) we have:
\begin{align}
 &\Psi \left( {t,{\bf{x}}} \right) = \Phi \left( {t,{\bf{x}}} \right) = \zeta \left( {\bf{x}} \right)\left[ {\frac{H}{a}\int\limits_{t_0 }^t {a\left( \tau  \right)d\tau  - 1} } \right], \\& 
 \delta \rho \left( {t,{\bf{x}}} \right) =  - \zeta \left( {\bf{x}} \right)\frac{{\mathop {\bar \rho}\limits^ .  }}{a}\int\limits_{t_0 }^t {a\left( \tau  \right)d\tau } , \\ &
 \delta p\left( {t,{\bf{x}}} \right) =  - \zeta \left( {\bf{x}} \right)\frac{{\mathop {\bar p}\limits^ . }}{a}\int\limits_{t_0 }^t {a\left( \tau  \right)d\tau } , \\ &
 \delta u\left( {t,{\bf{x}}} \right) = \frac{{\zeta \left( {\bf{x}} \right)}}{a}\int\limits_{t_0 }^t {a\left( \tau  \right)d\tau } .
 \end{align}
On the other hand, if we take $ \zeta=0 $ Eq.(\ref{c7}) gives
\begin{equation}\label{c9}
\epsilon _0 \left( {t,{\bf{x}}} \right) =  - \frac{{\chi \left( {\bf{x}} \right)}}{a},
\end{equation}
where $ \chi(\textbf{x}) $ is an arbitrary function on the $ S^3(a) $. Note that in this case $ \eta_i $ is a\textit{ Killing vector} of $ S^3 $. By substituting Eq.(\ref{c9}) in Eqs.(\ref{b9})-(\ref{c4}) we derived the second set of solutions as follows
\begin{align}
 &\Psi \left( {t,{\bf{x}}} \right) = \Phi \left( {t,{\bf{x}}} \right) = \chi \left( {\bf{x}} \right)\frac{H}{a} , \\ &
 \delta \rho \left( {t,{\bf{x}}} \right) =  - \chi \left( {\bf{x}} \right)\frac{{\mathop {\bar \rho}\limits^ .  }}{a} , \\& 
 \delta p\left( {t,{\bf{x}}} \right) =  - \chi \left( {\bf{x}} \right)\frac{{\mathop {\bar p}\limits^ . }}{a} , \\ &
 \delta u\left( {t,{\bf{x}}} \right) = \frac{{\chi \left( {\bf{x}} \right)}}{a}.
 \end{align}
Unlike the first solution, this solution is a decaying mode, so it can be neglected at late times and its existence is significant just for  counting of adiabatic solutions. In both solutions $ \frac{{\delta \rho \left( {t,{\bf{x}}} \right)}}{{\mathop {\bar \rho}\limits^ .  }} = \frac{{\delta p\left( {t,{\bf{x}}} \right)}}{{\mathop {\bar p}\limits^ . }} $ which means they are adiabatic solutions.\\
 In general, $ S^3 $ has four independent gradient conformal Killing vectors and six independent Killing vectors, however, we have totally two independent solutions for perturbations equations in super-Hubble scales.\\
 It can be shown that whatever would happen during the inflation, if the universe subsequently spends sufficient time in a state of local thermal equilibrium with conserved quantities, then the perturbations become adiabatic and they remain adiabatic, even when the conditions of local thermal equilibrium are no longer satisfied\cite{Ref44}.
 
\section{Conclusion and summary}
\label{sec6}
The de Sitter background is maximally extended and also maximally symmetric if only if $ K=1 $ i.e. its spatial section is closed. For this purpose, we obtained the required linear perturbation field equations and then proved the existence of two independent adiabatic solutions for these equations in the time interval when perturbations scales go outside of the Hubble horizon. We showed the curvature perturbation on the uniform density slices in a spatially closed universe is proportional to the divergence of the conformal Killing vector of $ S^3 $. This indicates some perturbative cosmological potentials in the time intervals when the scales of the majority of perturbative modes become longer than the Hubble horizon, reduce to the geometrical properties of the background. In comparison with the adiabatic solutions in the spatially flat background, it seems the curvature has no direct role when $ aH\gg1 $ , but dependence of $ \zeta(\textbf{x}) $  to the  background geometry manifests even outside the horizon the curvature is significant. We also investigate stochastic properties of the perturbation fields in a spatially closed background and show that the spectrums of them are discrete due to the compactness of $ S^3(a) $.

\end{document}